# Anomalous lattice dynamics in a σ-$Fe_{60}V_{40}$ alloy: Mossbauer spectroscopic study

Stanisław M. Dubiel[1*] and Jan Żukrowski[2]

AGH University of Science and Technology, [1]Faculty of Physics and Applied Computer Science, [2]Academic Center for Materials and Nanotechnology,

al. A. Mickiewicz 30, 30-059 Krakow, Poland

**Abstract**

Lattice dynamics in a σ-$Fe_{60}V_{40}$ compound, which shows a re-entrant magnetism and orders ferromagnetic ally at $T_C \approx 170$K, was investigated with the Mossbauer spectroscopy in the temperature interval of 5-300 K. Two relevant spectral parameters viz. the average center shift, <CS>, and the relative recoil-free fraction, $f/f_o$, were explored. The former yielded the Debye temperature, $T_{D1}$, and the mean-square velocity of vibration, <$v^2$>, while the latter $T_{D2}$ and the mean-square amplitude of vibrations, <$x^2$>. Significant differences in the lattice-dynamical behaviors in the magnetic and paramagnetic phases were revealed. In particular, the values of $T_D$ were notably lower and those of $f/f_o$ greatly higher in the former. This anomalous result has likely its origin in a remarkably high inharmonic contribution to the vibrations found for the ground magnetic state (spin-glass). Especially anomalous behavior vs. temperature exhibits <$x^2$> where four well defined ranges could have been identified and ascribed to the paramagnetic, ferromagnetic and two spin-glass phases. Linear correlations between <$v^2$>-<$x^2$> were found within each of the four ranges. They enabled determination of force constants, hence a change of the potential energy, $E_p$, in each of the ranges. The total change of $E_p \approx 30$ meV while the corresponding one of the kinetic energy, determined from the knowledge of <$v^2$>, was $E_k \approx 21$ meV. The lack of balance between $E_p$ and $E_k$ follows from the anharmonic lattice-dynamical behavior observed in the spin-glass state. The results give a strong evidence that magnetism can significantly affect the lattice dynamics.

[*] Corresponding author: Stanislaw.Dubiel@fis.agh.edu.pl (S. M. Dubiel)

## I. Introduction

A thorough knowledge and good understanding of atomic lattice vibrations in solids, are essential for the proper understanding of their physical properties such as thermal conductivity, heat capacity, vibrational entropy, Debye temperature, electron-phonon coupling as well as the noise of electronic devices. One of open questions in the field is a possible relationship between magnetism and the lattice vibrations. A contribution of an electron-phonon interaction (EPI) to magnetization of metallic systems is expected to be small, as, in general $E_D/E_F \leq 10^{-2}$ [1], where $E_D$ is the Debye energy and $E_F$ is the Fermi one. Consequently, the effect of magnetism on the lattice dynamics in such systems is expected to be negligible. However, following Kim [1] the effect of the electron-phonon coupling can be strongly enhanced below the Curie temperature, $T_C$, in an itinerant ferromagnet. Very good candidates for verifying these predictions are σ-phase Fe-X (X=Cr, V, Re, Mo) alloys because they exhibit a highly itinerant type of magnetism [2]. The σ-phase can occur in binary and ternary alloy systems with the common crystallographic structure ($D^{14}_{4h}$-P4$_2$/mnm), while its physical properties depend, in general, on the system and its composition. Our previous measurements on quasi-equiatomic σ-phase Fe-Cr and Fe-V alloys gave evidence that quantities relevant to the lattice dynamics viz. SOD and the recoil-free fraction, $f$, [3] as well as the sound velocity [4] were significantly different in paramagnetic and magnetic phases.

The present study was carried out on a σ-Fe$_{60}$V$_{40}$ intermetallic alloy to further explore the issue and shed more light on it. The magnetic ordering (Curie) temperature of the alloy is ~170K [5] what makes it very suitable for studying the lattice dynamics in wide temperature ranges in which paramagnetic and magnetic phases exist. The results obtained give clear evidence that the lattice dynamics in the magnetic phase is very different than the one in the paramagnetic phase.

## II. Experimental

### A. Sample preparation and characterization

Master alloy of a nominal composition α-$Fe_{60}V_{40}$ was prepared by melting appropriate amounts of Fe (99.95%purity) and V (99.5%purity) in an arc furnace under protective argon atmosphere. A loss of mass caused by the melting corresponded in the concentration uncertainty ±0.1 at%. The ingot was flipped over and re melted few times before it was solution treated at 1273K for 72h. Finally, it was quenched onto a block of brass kept at 295K. The transformation into the σ-phase was performed by annealing the solution-treated ingot at T = 973 K for 14 days. The verification of the α-to-σ phase transformation was done by recording room-temperature X-ray and neutron diffraction patterns on powdered sample as described in detail elsewhere [6].

**B. Mössbauer spectra measurements and analysis**

Mössbauer spectra were recorded in a temperature (*T*) interval of 5-295 K using a standard spectrometer working in a constant acceleration mode and two cryostats: Janis Research 850-5 Mössbauer Refrigerator System in the range of 5-100 K, and in the Janis SVT-400 in the range 100-300 K. Temperature was stabilized to the accuracy < ±0.1 K. 14.4 keV γ-rays were supplied by a $^{57}Co(Rh)$ source. Examples of the spectra recorded in a para- and magnetic states of the sample are presented in Fig. 1 (left panel). A transmission integral method for the spectra analysis was used. All three hyperfine interactions were taken into account. Each spectrum was considered to be composed of five components due to the fact that Fe atoms are present on all five lattice sites in the unit cell of σ. The shape of each component was assumed to have the Voight's profile. Relative contributions of the components, $W_k$, were equal to the corresponding relative lattice site occupancies by Fe atoms as revealed by the neutron diffraction experiment [6]. They were kept constant in the fitting procedure. A center shift, $CS_k$, (k=1,2,3,4,5)) of each component was a sum of the isomer shift characteristic of the given sub lattice, as reported elsewhere [7], and a second-order Doppler shift (SOD) term. The latter was common to all five components and treated as free parameter to allow for the temperature effect. Its temperature dependence was used to determine the Debye temperature and the mean-square velocity of lattice vibrations – see below. Concerning the hyperfine field (*B*) each of the five components was assumed to have a Gaussian distribution. The average hyperfine field, <*B*>, was obtained by integrating the Gaussians. The left panel of Fig. 1 illustrates examples of the resulting hyperfine distribution curves.

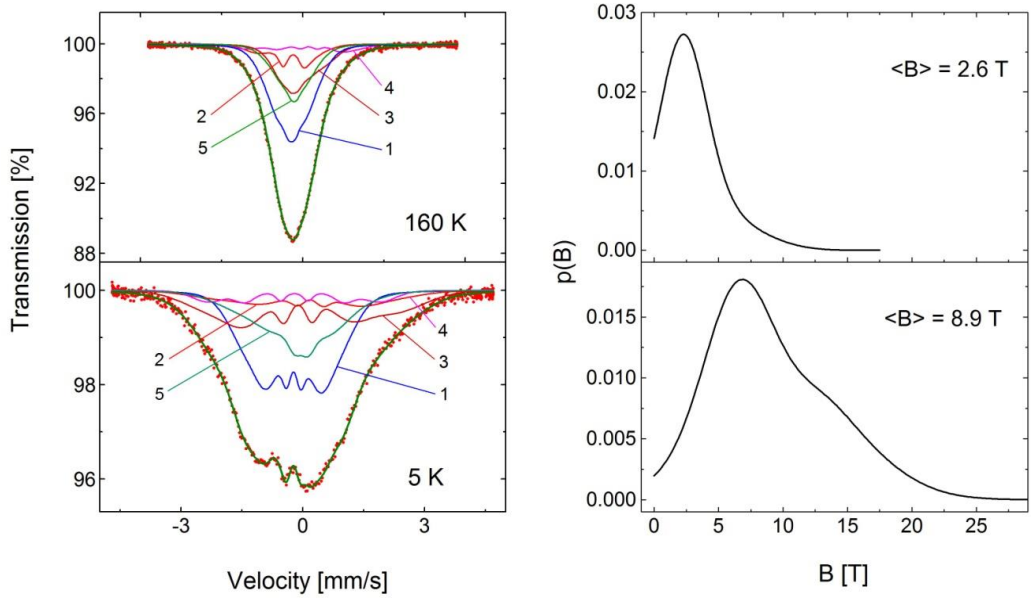

**Fig. 1**
Left panel: $^{57}$Fe Mössbauer spectra recorded on σ-Fe$_{60}$V$_{40}$ at 160 K (paramagnetic phase) and 5 K (magnetic phase). The five components corresponding to the five lattice sites are indicated. Right panel: Hyperfine distributions curves derived from the spectra shown in the left panel. The average values of the hyperfine field, <B>, are displayed.

### III. Results and discussion

#### A. Curie temperature

Temperature dependence of <B> is illustrated in Fig. 2. The data were fitted to the Brillouin function. It can be noticed that <B> increases anomalously below ~90K, a feature known to occur in re-entrant spin-glasses e. g. [8,9] to which also belongs the presently studied alloy [10]. Consequently, the <B>(T) data were analyzed in terms of two two Brillouin curves both with J=5/2. The increase of <B> in the low-temperature limit equals 0.7(1) T reflects freezing of the transverse component of spin [8,9]. The analysis of the data yielded 171.7(2) K for the Curie temperature a value that agrees well with the one determined with magnetization measurements [10].

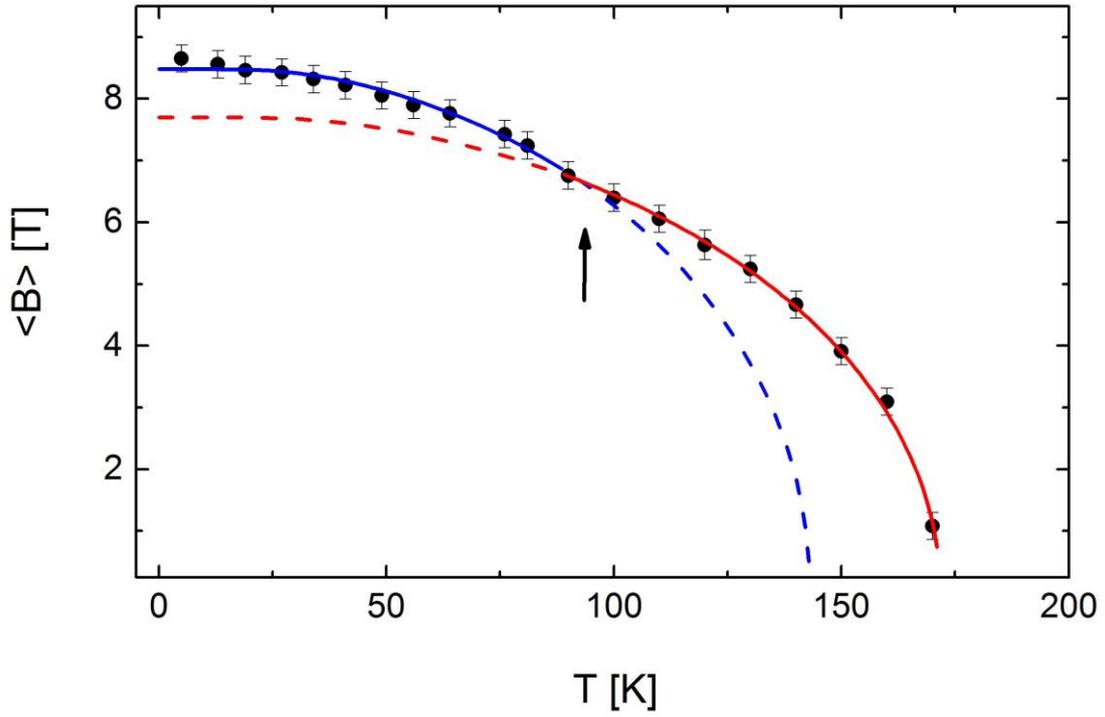

**Fig. 2**

Average hyperfine field, *<B>*, vs. temperature *T*. The data were analyzed in terms of two Brillouin functions due to an anomaly at ~90K indicated by arrow.

**B. Lattice dynamics**

**B1. Temperature dependence of the center shift**

A temperature dependence of the center shift, *CS(T)*, can be expressed by the following equation:

$$CS(T) = IS(0) - \frac{3k_B T}{2mc}\left[\frac{3T_D}{8T} + \left(\frac{T}{T_D}\right)^3 \int_0^{T_D/T} \frac{x^3}{e^x - 1}dx\right] \quad (1)$$

Where *IS(0)* stays for the isomer shift (temperature independent), $k_B$ is the Boltzmann constant, *m* is a mass of $^{57}$Fe atoms.

The second term in eq. (1), known as SOD depends on the mean-squared velocity of the vibrating atoms, <$v^2$>, via the following equation:

$$SOD = -\frac{E_\gamma}{2c^2}\langle v^2 \rangle \qquad (2)$$

Where $E_\gamma$ stands for the energy of γ-rays (here 14.4 keV) and $c$ is the velocity of light. The CS(T) dependence found in the present study is illustrated in Fig. 3. The best-fit of eq. (1) to the measured data, shown in Fig. 4 as a solid line, yielded $T_{D1}$ = 485(15) K for 170 ≤ T ≤ 300 K, and $T_{D1}$ = 322(17)K for 5 ≤ T ≤ 150 K. The difference in $T_D$ (designated in Table 1 as $T_{D2}$) for the paramagnetic and magnetic phases is unambiguous, and it proves that magnetic ordering really can affect the lattice dynamics.

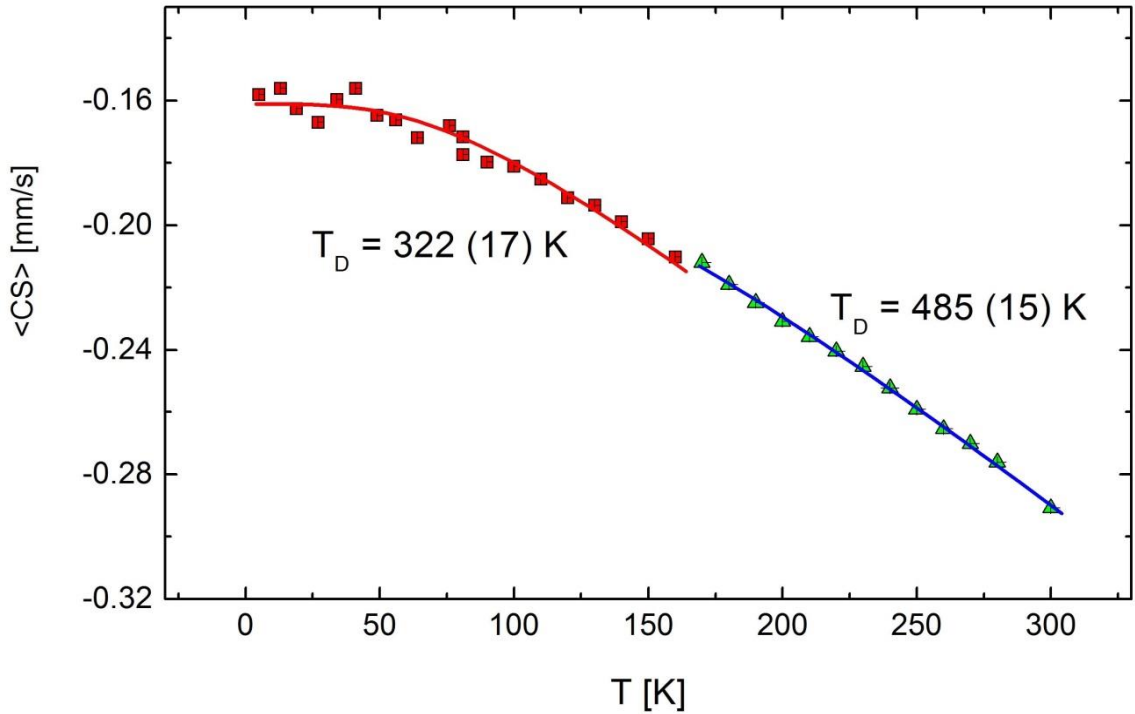

**Fig. 3**

Temperature dependence of the average center shift, <CS>. The best-fit curves in terms of eq. (1) are indicated for the paramagnetic and magnetic phases. Values of the Debye temperatures are displayed.

## B2. Temperature dependence of the mean-square velocity

The mean-squared velocity, <$v^2$>, can be calculated from eq. (2). Its temperature dependence is shown in Fig. 4.

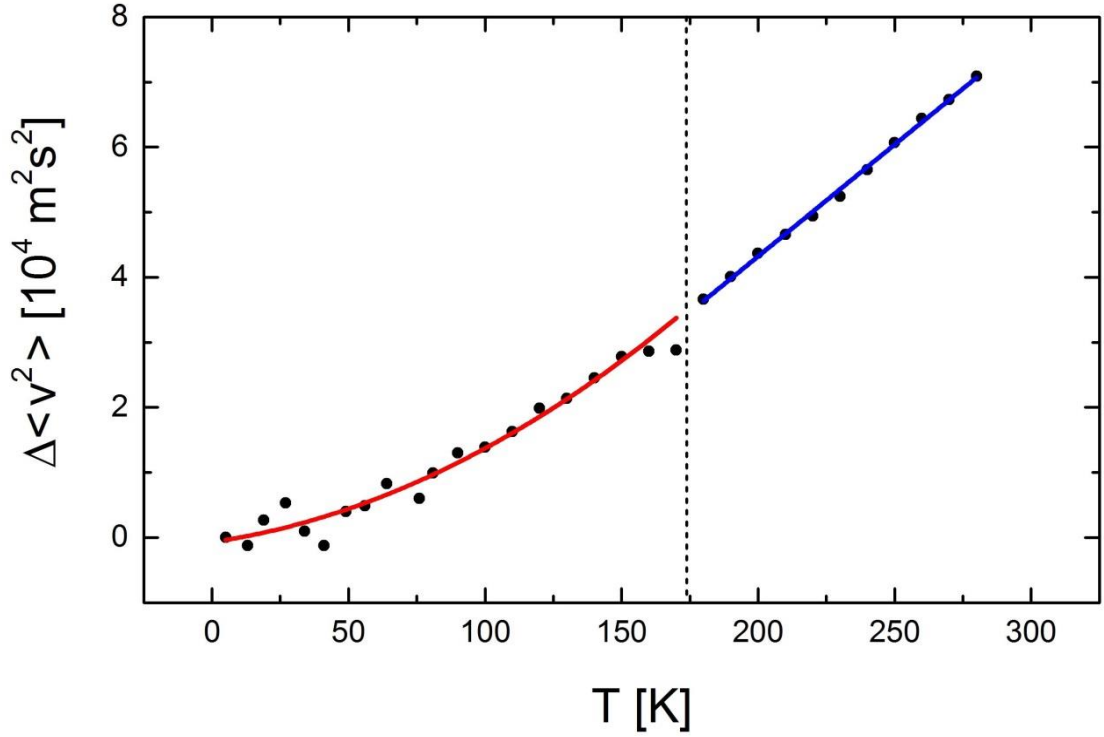

**Fig. 4**
Temperature dependence of the mean-square velocity. The lines show the best parabolic fits for the *T*-ranges in which the paramagnetic and magnetic phases exist. The approximate position of the Curie temperature is marked by the dashed line.

## B3. Temperature dependence of the *f*-factor

The recoil-free fraction, *f*, is defined by the following equation:

$$f = \exp[-(\frac{E_\gamma}{\hbar c})^2 \langle x^2 \rangle] \quad (3)$$

Where <$x^2$> is the mean-square amplitude of vibrating atoms. It is related to θ$_D$ via the following expression:

$$f = \exp[-\frac{6E_R}{k_B T_D}\{\frac{1}{4}+(\frac{T}{T_D})^2 \int_0^{T_D/T}\frac{x}{e^x-1}dx]  \qquad (4)$$

Where $E_R = \frac{E_\gamma^2}{2mc^2}$ is the recoil energy.

In the approximation of a thin absorber, the *f*-factor is proportional to a spectral area, *A*. In practice one uses a normalized spectral area, $A/A_o$, as a measure of the relative *f*-factor, $f/f_o$ ($A_o$ being the spectral area at the lowest temperature – 5 K in this case). The temperature dependence of $ln(f/f_o)$ is presented in Fig. 5.

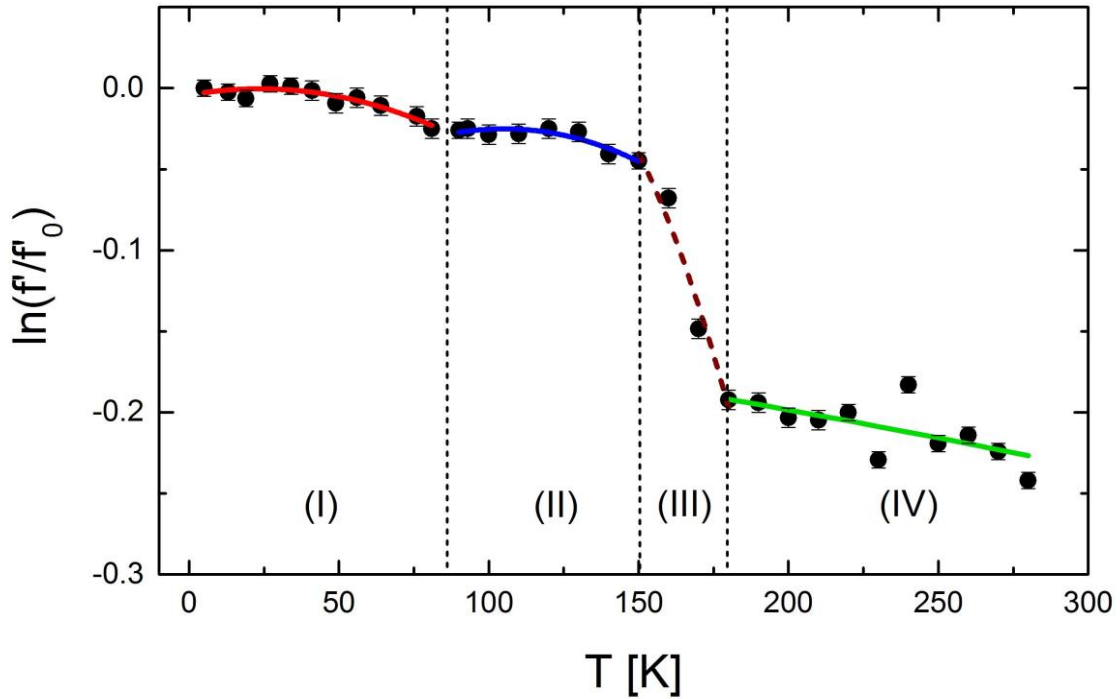

**Fig. 5**

Ln($f/f_o$) vs. temperature, *T*. Five ranges, separated by dashed lines, can be identify: I and II in which the behavior is not linear, III and IV in which the behavior is linear. The lines stand for the best-fits to the data.

The data shown in Fig. 5 evidently show that the relationship is unusual. First of all, a step-like increase is observed on going from the paramagnetic to magnetic phases. It is indicative of a lattice hardening in the magnetic phase. The overall relation can be divided into four ranges: I and II in which the behavior is non-linear, and III and IV with a linear dependence. The plot reflects a re-entrant character of magnetism revealed in this compound [10]. Namely, I and II cover the temperature range in which a spin-glass occurs, III the one with a ferromagnetic ordering and IV coincides with the paramagnetic phase. The non-linear behavior of ln$f$ indicates anharmonic vibrations. Their effect on $f$ can be expressed as follows [11,12]:

$$\ln f = -\frac{6E_R T}{T_D^2}(1+\varepsilon T+...) \qquad (5)$$

The data displayed in Fig. 5 analyzed with eq. (5) yielded for $T_D$ (designated in Table 1 as $T_{D1}$) the following values: 657(150) K, 248(30) K, 104(20) K, 577(43) K for the ranges I, II, III and IV, respectively. Noteworthy, the average value over I, II and III (magnetic phase) amounts to 343(31) K which is about the same as determined from the temperature dependence of the average center shift for the magnetic phase.

The anharmonic coefficient, $\varepsilon$, being $-2.3 \cdot 10^{-2}$ K$^{-1}$ for the range I and $-4.6 \cdot 10^{-3}$ K$^{-1}$ for the range II. Noteworthy, these values of $\varepsilon$ are very high e. g. the former is by a factor of 10 larger than the one determined for Fe impurities embedded into Cr matrix [13].

### B4. Temperature dependence of the mean-square amplitude

The mean-square amplitude of vibrations, $<x^2>$, can be determined via eq. (3). Its temperature dependence, relative to the one at 5 K, $\Delta<x^2>$, is illustrated in Fig. 6. It illustratively shows that the vibrations do not change monotonically with temperature. They hardly depend on temperature in the ranges I and II (spin-glass) while a steep decrease occurs in the range III where ferromagnetic ordering exists. An intermediate behavior takes place in the paramagnetic phase (range IV).

### B5. Force constant

The force constant, $D$, can be determined based on a linear correlation between $<x^2>$ and $<v^2>$, as reported elsewhere [14]. The corresponding relation - displayed in Fig. 7 - is not linear in the present case. However, to a good approximation, the behaviors within the four ranges are linear – see Fig. 8. The latter can be thus used to determine $D = m \cdot \alpha$, where $m$ is a mass of a vibrating atom ($^{57}$Fe) and $\alpha$ stands for the slope of the best-fit line. The $D$-values obtained in this way are displayed in Table 1. It follows that the hardest coupling experience Fe atoms in the SG1 phase ($D = 4510$ N/m), next in the SG2 phase ($D=667$ N/m) and the weakest coupling they sense when present in the FM phase. In other words, a significant decoupling of Fe atoms occurs in the latter.

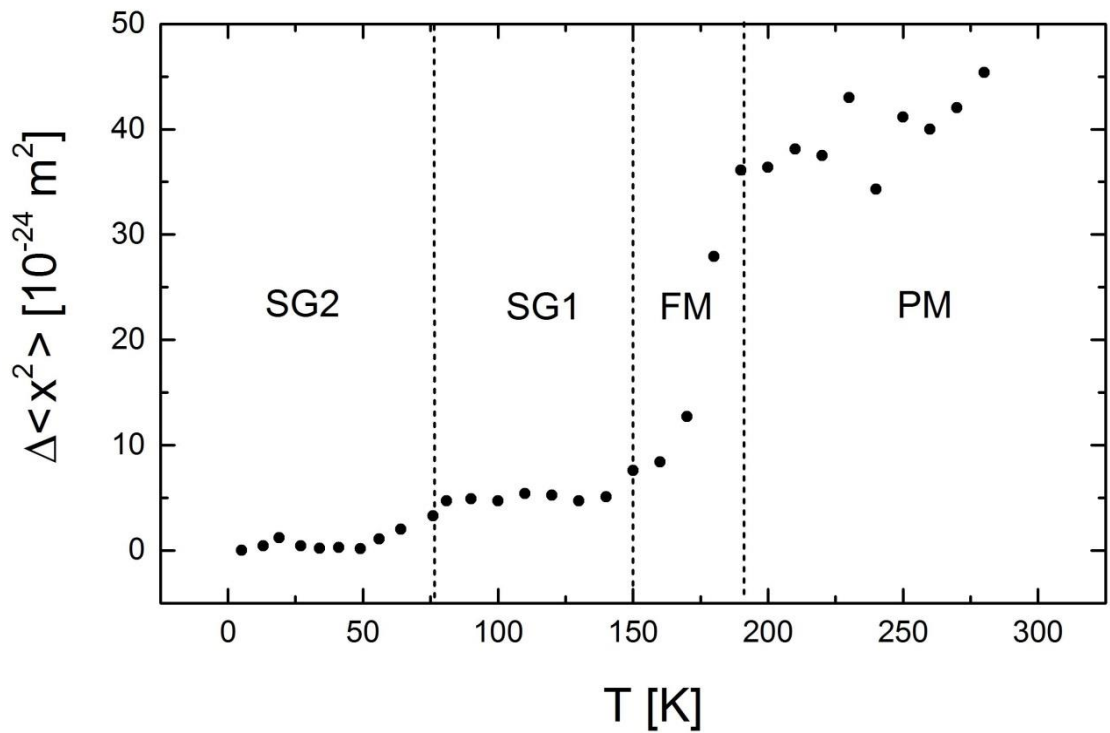

**Fig. 6**

Temperature dependence of the relative mean squared amplitude of vibrations, $\Delta<x^2>$. Vertical lines indicate four ranges into which the relation has been divided. Corresponding phases are indicated.

**Table 1**

Values of the Debye temperatures, $T_{D1}$ and $T_{D2}$, force constant, $D$, anharmonicity parameter, $\varepsilon$, and change of the potential energy, $\Delta E_p$, as determined for the four temperature ranges (phases).

| $T$ [K] | Phase | $T_{D2}$ [K] | $T_{D1}$ [K] | $D$ (N/m) | $\varepsilon$ [$10^{-3}$] | $\Delta E_p$ [meV] |
|---|---|---|---|---|---|---|
| 5-80 | SG2 | 657(140) | 322(17) | 667 | -23 | 9.8 (0→9.8) |
| 80-140 | SG1 | 248(30) | | 4510 | -4.6 | 7.0 (9.8→16.8) |
| 140-170 | FM | 104 (60) | | 42 | 0 | 3.9 (16.8→20.7) |
| 170-280 | PM | 577(43) | 485(15) | 411 | 0 | 9.0 (20.7→29.7) |

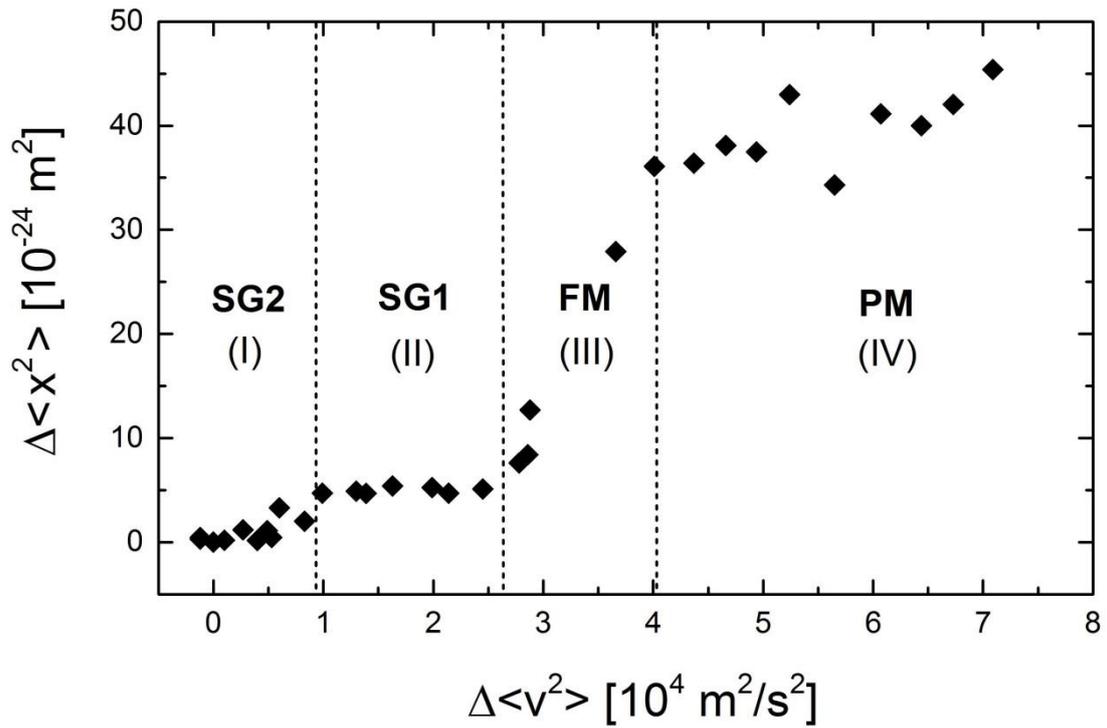

**Fig. 7**

Relationship between the relative squared amplitude of vibrations, $\Delta\langle x^2\rangle$, and the corresponding relative squared velocity of vibrations, $\Delta\langle v^2\rangle$. Vertical lines indicate the division into four ranges.

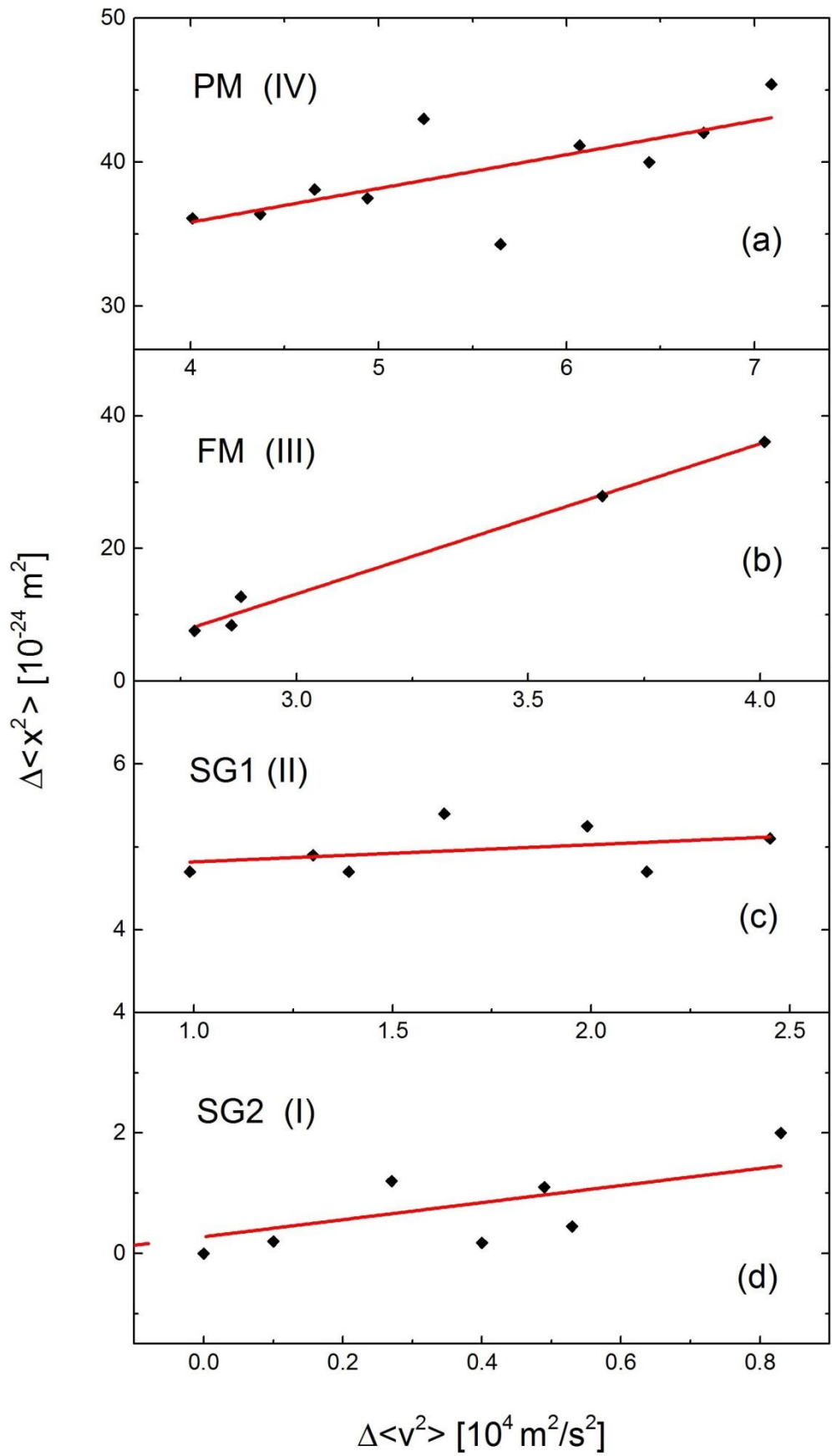

**Fig. 8**

Mean-square amplitudes of vibrations, $<x^2>$, vs. mean-square velocity of vibrations for the four ranges (phases). The best fit linear fits to the data are shown, too.

**C. Energy relations**

The knowledge of $f/f_o$ enabled, via eq. (3), determination of a relative change of the mean-square amplitude of vibrating atoms, $\Delta<x^2>=<x^2>-<x_o^2>$, and, in turn, a relative temperature-induced change of the potential energy, $\Delta E_p=0.5 \cdot D \cdot \Delta<x^2>$, within the particular ranges viz. I, II, III and IV. The corresponding changes, both absolute and relative, are displayed in Table 1. As the relative changes took place in different ranges of temperature, one should rather calculate a change per Kelvin, $\delta E=\Delta E_p/\Delta T$, for the given temperature interval to figure out whether or not there are differences between the phases. The values of $\delta E$ are: 0.13, 0.12, 0.13 and 0.10 meV/K for SG2, SG1, FM and PM phases, respectively. This means that in the magnetic phase (SG+FM), a change of temperature by 1K changes the potential energy of Fe atom vibrations by ~30% more than in the paramagnetic phase. The total change in $E_p$ that took place in the temperature interval between 5 and 280 K is equal to 29.7 meV.

In turn, the mean-square velocity of the vibrating Fe atoms, $<v^2>$, hence that of the kinetic energy, $E_k=0.5 \cdot m \cdot <v^2>$, could be determined via eq. (2). As we intend to compare it with the potential energy of the vibrations which, in the present experiment, can be determined only relatively – see the previous paragraph – we have to determine a change of the kinetic energy, $\Delta E_k=0.5 \cdot m \cdot (<v^2>-<v_o^2>)$, relative to its value at the lowest measured temperature, $E_{ok}=0.5 \cdot m \cdot <v_o^2>$. The value of $\Delta E_k$ =20.8 meV which is close to a change of the thermal energy, $\Delta E=k_B \Delta T$=23.6 meV, but significantly less than the corresponding change of the potential energy. This difference can be understood in terms of the anharmonic vibrations ($\varepsilon \neq 0$) found in the spin-glass state ($T$=5-140 K). In the temperature range of 140-280 K $\Delta E_p$=12.9 meV and $\Delta E_k$=13.4 meV i.e. $\Delta E_p=\Delta E_k$ with the accuracy of ±4%. This corroborates with the fact that in the FM+PM phases the vibrations are harmonic ($\varepsilon$=0),

**IV. Conclusions**

Based on the results presented in this paper a general conclusion relevant to Fe atom lattice dynamics can be drawn, namely the dynamics in the magnetic phase is significantly different than the one in the paramagnetic phase. In particular:

- The Debye temperature determined from the temperature dependence of the center shift is equal to 485(15) K for the paramagnetic phase and to 322(17) K for the magnetic one.

- The Debye temperature determined from the temperature dependence of the recoil-free fraction is equal to 577(15) K for the paramagnetic phase and to 343(31) K for the magnetic one (the latter being an average over three magnetic sub phases).

- The recoil-free fraction, hence the mean-square amplitude of vibrations, shows a step-like behavior vs. temperature with four ranges associated with four phases viz. paramagnetic (PM), ferromagnetic (FM), and two spin-glass (SG1, SG2) in the system.

- The linear correlations between the mean-square velocity and the mean-square amplitude of vibrations that hold within the four ranges permitted determination of the force constants, *D*, viz. 411, 42, 4510 and 667 N/m for PM, FM, SG1 and SG2, respectively.

- The change of the potential energy, $\Delta E_p$, in the whole temperature range (5-280 K) is by ~50% higher than the corresponding change of the kinetic energy, $\Delta E_k$, which testifies to anharmonic lattice vibrations in the studied sample.

- The anharmonicity of the lattice vibrations occurs in the spin-glass phase, the stronger one being in the SG2 sub phase.